\documentclass[twocolumn,aps]{revtex4}
\usepackage[english]{babel}
\usepackage{graphicx}
\usepackage{epsf}
\usepackage{dcolumn}

\begin{document}
\title{ On the density of states for the Hubbard model:
pseudo-particle Keldysh diagram method - an alternative to DMFT?}
\author{ P.I. Arseyev , N.S. Maslova $^\dag$\thanks{e-mail: nsmaslova@gmail.com}, \\
 P.N.Lebedev Physical Institute of RAS, Leninskii pr.53, 119991 Moscow, Russia \\
 $^\dag$ Department of Physics, Moscow State University, 119992 Moscow, Russia}
\date{\today }
\begin{abstract}
It is shown how to construct Keldysh diagram technique for  pseudo - particle approach to the Hubbard model.
We propose  self consistent equations for
pseudo particle and electron Green functions in Keldysh diagram technique.
Nonlocal effects (spatial dispersion) are included in single impurity problem in this method.
Thus we can get rid of the artificial central peak (of Kondo type) in the density of states which is
inevitable in Dynamical Mean Field Theory (DMFT).
The changes in the density of states for 2D Hubbard model
due to variation of Coulomb repulsion $U$ and electron
concentration are analyzed.
\end{abstract}

\pacs{71.10.Fd, 71.27.+a}

\maketitle

Materials with correlated electrons demonstrate a great variety of unusual interesting phenomena.
But up to now the correct theoretical description of their properties encounters with great difficulties.
One of the main methods used nowdays is the so called DMFT
(dynamical mean field theory)\cite{Kotl1,Kotl2}. In spite of wide popularity of this method, it has
some considerable shortcomings. DMFT is formulated in the framework of temperature diagram technique,
so one needs to make analytical continuation procedure to calculate density of states; a non-physical artifact
peak in the density of states appears inside dielectric gap for the Hubbard model; computations are very
cumbersome and require much time.

One of the alternative methods is based on introducing non-physical pseudo-particles ("slave-particles")
 to describe
independently each state in correlated system \cite{Abricos,Barnes,Col}. This method requires exact
constraint on pseudo-particle numbers at each site which results in strong modification of usual
diagram technique.
To avoid difficulties of this modified diagram technique some authors used mean field approximation in
functional integral formulation of this method \cite{Barnes,Col,Kotl2}. But the validity of obtained results
is a matter of question.
An attempt to generalize pseudo-particle method for nonequilibrium Anderson model was made in papers
\cite{Avi,Paaske1,Paaske2}. These authors obtained results only for a few lowest order diagrams
or for non-crossing approximation. The general rules for constructing diagram series in Keldysh
technique for pseudo particles method are absent.

In the present paper we show how to construct Keldysh diagram technique \cite{kel}
for the pseudo - particle (PP) approach
to the Hubbard model. Understanding of the general rules of PP diagram technique allowed us to suggest
self-consistent scheme of PP line calculations and to obtain reasonable results for electron density of states
for the Hubbard model with different Coulomb repulsion values and arbitrary electron
concentration.

Hamiltonian of the Hubbard model has the well known form:
\begin{equation}
\hat H = \sum\limits_{ij\sigma } {t_{ij} } c_{i\sigma }^ +  c_{j\sigma }
+ \sum\limits_{i\sigma } U n_{i\sigma } n_{i - \sigma }  +
\sum\limits_{i\sigma } ( \varepsilon _i  - \mu )c_{i\sigma }^ +  c_{i\sigma }
\end{equation}
where $c_{i\sigma }^ +$ is electron creation operator, $\varepsilon _i$ - on site electron energy, $\mu$ - chemical potential,
 $t_{ij}$ - hopping matrix element and $U$ - on site Coulomb
repulsion. One can introduce non-physical particles each of them is assigned to a definite
single site state \cite{Abricos, Barnes}. Creation operators of these pseudo particles (PP)
correspond to appearance of the following physical states:
\begin{equation}
 b^ +   \Rightarrow \,|0 > \qquad f_\sigma ^ +   \Rightarrow c_\sigma ^ +  |0 > \qquad d^ +
  \Rightarrow c_ \uparrow ^ +  c_ \downarrow ^ +  |0 >
\end{equation}
$b$ and $d$ are bose and $f$ - fermi PP.
Unphysical states are eliminated by the constraint for each site:
\begin{equation}
    \label{constr}
 \hat N_0  = \sum\limits_\sigma  {} f_\sigma ^ +  f_\sigma   + b^ +  b + d^ +  d = 1
\end{equation}
Creation operator of a physical electron is expressed as:
\begin{equation}
 c_\sigma ^ +   = f_\sigma ^ +  b + d^ +  f_{ - \sigma }
\end{equation}
In this PP representation the on-site Hamiltonian with Coulomb interaction between electrons
looks like a Hamiltonian for non-interacting PP:
\begin{equation}
 \hat H^0  = \sum\limits_{i} \left( \sum\limits_{\sigma}\varepsilon  f_{i\sigma }^ +  f_{i\sigma }  +
 (2\varepsilon  + U)\,d_i ^ +  d_i  + 0 \bullet b_i^ +  b_i \right)
\end{equation}
From now on all single electron energies are measured from the chemical potential $\mu$.
Hopping between the sites now looks like interaction between pseudo particles:
\begin{equation}
  \label{tij}
\hat H_{int}  =
\sum\limits_{ij\sigma } {t_{ij} } (f_{i\sigma }^ +  b_i  +
d_i ^ +  f_{i - \sigma } )(f_{j\sigma }^{} b_j ^ +   + d_j ^{} f_{j - \sigma } ^ +  )
\end{equation}
Any physical state should contain only one pseudo particle. In this  subspace
determined by the constraint (\ref{constr})  the mapping is exact.
The projection to this pseudo particle subspace can be done by the following trick \cite{Abricos}.
We add some large positive constants $\lambda_i$ to PP energies at all sites:
\small
\begin{equation}
\hat H_\lambda ^0  = \sum\limits_i \left[\sum\limits_\sigma  ( \varepsilon  + \lambda_i )f_\sigma ^ +  f_\sigma
+ \lambda_i b^ +  b + (2\varepsilon  + U + \lambda_i )d^ +  d \right]
\end{equation}
\normalsize
In the present paper we consider thermodynamics averages as initial basic elements
for Keldysh diagram technique.
Then states with $k$ PP on site "$i$" have weight $e^{ - k\lambda_i /T}$ for large $\lambda_i$
($T$ is the temperature).
So only single PP states on any site "$i$" can be retained in any average $< ....... >$
by the following operation:
\begin{equation}
\mathop {\lim }\limits_{\lambda_i  \to \infty } \{ \,e^{\lambda_i /T}  \times  < ....... > \}
\end{equation}
States with two or more PP have exponentially small weights and vanish in the limit
$\lambda_i  \to \infty$. The unphysical "vacuum state" with no PP is excluded because the
Hamiltonian and any physical operator are normally ordered combinations of PP operators.
After this operation PP occupation numbers for each site are determined as:
\begin{equation}
  \label{n_0}
\begin{array}{l}
 n_0  = (Z_0 )^{ - 1} \mathop {\lim }\limits_{\lambda  \to \infty } e^{\lambda /T} n_\lambda
 = \frac{{e^{ - \varepsilon /T} }}{{1 + 2e^{ - \varepsilon /T}  + e^{ - (2\varepsilon  + U)/T} }}\: \\
 b_0  = (Z_0 )^{ - 1} \mathop {\lim }\limits_{\lambda  \to \infty } e^{\lambda /T} b_\lambda
 = \frac{1}{{1 + 2e^{ - \varepsilon /T}  + e^{ - (2\varepsilon  + U)/T} }} \\
 d_0  = (Z_0 )^{ - 1} \mathop {\lim }\limits_{\lambda  \to \infty } e^{\lambda /T} d_\lambda
  = \frac{{e^{ - (2\varepsilon  + U)/T} }}{{1 + 2e^{ - \varepsilon /T}  + e^{ - (2\varepsilon  + U)/T} }} \\
 \end{array}
\end{equation}
where
\begin{equation}
Z_0  = Sp_{N=1} \left( {e^{ - \beta \hat H} } \right) =
\mathop {\lim }\limits_{\lambda  \to \infty } e^{\beta \lambda }
Sp\left[ {e^{ - \beta \hat H_\lambda  } \widehat{N_0 }} \right]
\end{equation}
This PP ocupation numbers satisfy the required constraint:
\begin{equation}
{\rm{b}}_0 {\rm{ + 2n}}_0 {\rm{ + d}}_0 {\rm{ = 1}}
\end{equation}
Electron spectrum and density of states can be obtained from usual Green functions:
\begin{equation}
G_{\sigma ij}^{ \alpha,\beta } (t,t') =  - i < T_c c_{i\sigma}  (t),c_{j\sigma} ^ +  (t') >
\end{equation}
where $T_c$ means ordering on the Keldysh contour \cite{kel}.
Single electron Green functions looks like two particle objects in PP representation, for example:
\small
\begin{equation}
\begin{array}{l}
 G_{\sigma ij}^{ -  - } (t,t') = \\
 = - i < Tb^+_i f_{i\sigma } (t),f_{j\sigma } ^ +  b_j(t') >
 - i < Tf_{ - \sigma i}^ +  d(t)_i,d_j^+  f_{ - \sigma j} (t') >  \\
 - i < Tf_{ - \sigma i }^ +  d(t)_i,f_{\sigma j} ^ +  b_j(t') >
 - i < Tb_i^ +  f_{\sigma i} (t),d_j^+  f_{ - \sigma j} (t') >  \\
 \end{array}
 \end{equation}
 \normalsize
Before taking the limit $\lambda \to \infty$ the usual diagram rules are valid.
All diagrams include PP Green functions as if they were real particles. For example "lesser" PP
Green functions are
\begin{equation}
\begin{array}{l}
   \label {G_f}
 G_{f\sigma }^ <  (t - t') = in_f^\lambda  e^{ - i(\varepsilon _{}  + \lambda )(t - t')}  \\
 G_b^ <  (t - t') =  - ib^\lambda  e^{ - i\lambda (t - t')}  \\
 G_d^ <  (t - t') =  - id_{}^\lambda  e^{ - i(2\varepsilon  + U + \lambda )(t - t')}  \\
 \end{array}
\end{equation}
Retarded on-site electron Green function $G^R_{ii}$ without intersite transitions is a sum of two simple
closed loops (polarization operators) in PP representation (Fig.1).

\begin{figure}[h!]
\centerline{\includegraphics[width=6cm]{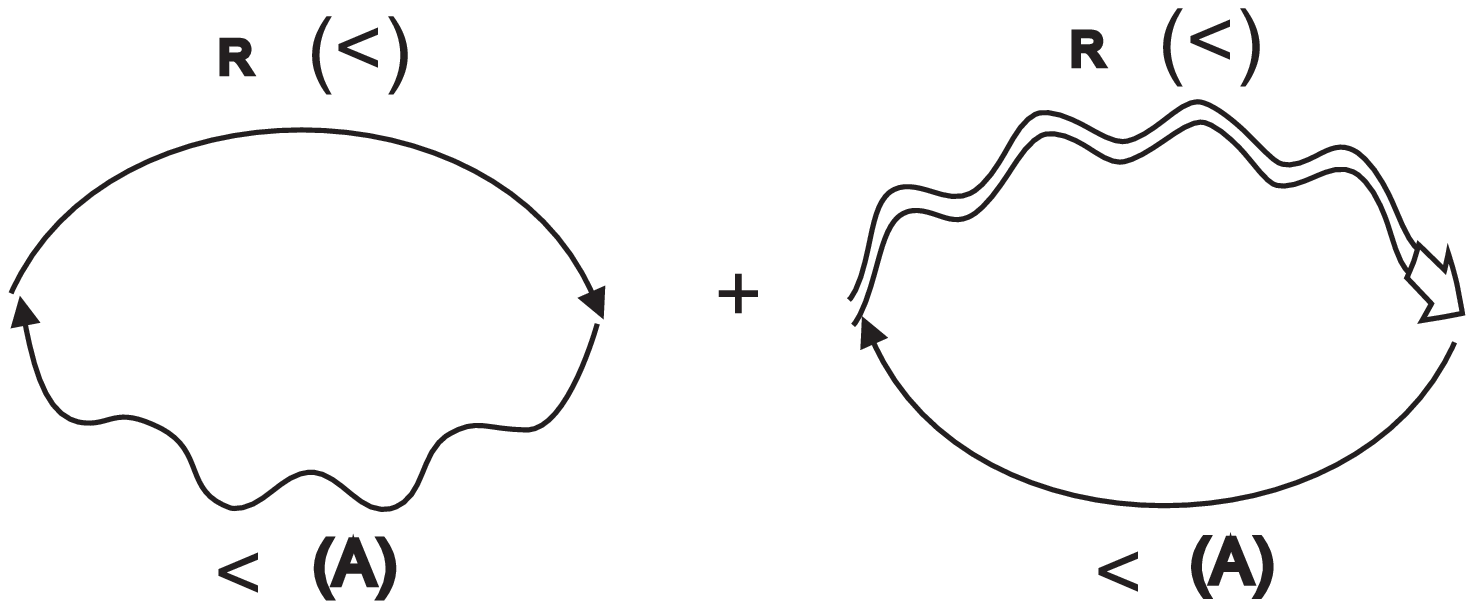}}
\caption{
Retarded on-site electron Green function $G^R_{ii}$. Solid line corresponds to pseudo-fermion function,
wavy line - to empty-site pseudo-boson and double wavy line - to double-occupied site pseudo-boson.
 }
\end{figure}

After projection to the physical subspace described above we have:
\begin{equation}
   \label {GR}
\begin{array}{l}
 G^R_{ii\sigma 0}(t - t') = \\
 i\int {\frac{{d\omega _1 }}{{2\pi }}} \left[ {B_{0}^< (t' - t)N^R_{\sigma 0}(t - t')
 + B_{0}^A (t' - t)N^<_{\sigma 0} (t - t')} \right. -  \\
 \left.  - N_{ - \sigma 0 }^ < (t' - t)D^R_0 (t - t')
  - N_{ - \sigma }^A (t' - t)D^ <  (t - t') \right] \\
 \end{array}
\end{equation}
Where PP Green functions $B,N,D$ appear instead initial PP functions $G_b,G_f,G_d$  after the projection
procedure:
\begin{equation}
\begin{array}{l}
 N_{0\sigma }^ < (t - t') = i n_0  e^{ - i \varepsilon (t - t')}  \qquad
 B^<_0  (t - t') =  - i b_0 \\
 D^<_0  (t - t') =  - i d_{0} e^{ - i(2\varepsilon  + U )(t - t')}
 \end{array}
\end{equation}
and $n_0,b_0,d_0$ are given by Eq. (\ref{n_0}). Retarded PP functions are:
\begin{equation}
  \label{N_0}
\begin{array}{l}
 N_{0\sigma }^R (t - t') = -i \theta(t - t') e^{ - i \varepsilon (t - t')}
 B^R_0  (t - t') =  - i \theta(t - t') \\
 D^R_0  (t - t') =  - i \theta(t - t') e^{ - i(2\varepsilon  + U )(t - t')}  \\
 \end{array}
\end{equation}

After Fourier transformation we obtain simple Green function for the single-site Hubbard model:
\begin{equation}
   \label{G0}
G _0^R(\omega)  = \frac{n_0  + b_0 }{\omega  - \varepsilon  + i\delta } +
\frac{{n_0  + d_0 }}{{\omega  - \varepsilon  - U + i\delta }}
\end{equation}

Intersite transitions (\ref{tij}) correspond to four types of the two particle vertexes
connecting closed loops for
neighboring sites. So any diagram consists of some number of closed single-site PP loops
connected with intersite hopping lines $t_{ij}$. Before the projection procedure is made
any PP "lesser" Green function
(\ref{G_f}) is proportional to ${\rm exp}(-\lambda_i/T)$. So after taking the limit $\lambda_i\to \infty$ only
diagrams with one "lesser" Green PP function at a given site are retained. This fact gives rise to the following
rules for constructing the diagrams with full account of the constraint on the PP total number:

1) Only one pseudo - particle loop for any site can appear in a diagram;
2) Only one PP "lesser" function in any loop can be present. It is substituted by
renormalized PP occupation number (\ref{n_0});
3) Only R and A parts of any other PP Green functions are retained in any on-site loop;
4) Oscillating multipliers ${\rm exp}(-i\lambda_i t)$ are cancelled in any vertex and should be omitted.

We can construct perturbation series in intersite hopping $t_{ij}$.
First order diagrams are proportional to $t_{ij}^2/\varepsilon^2$ or $t_{ij}^2/(\varepsilon+U)^2$.
These diagrams consist of two PP loops for neighboring sites connected with two hopping vertexes.
One of these diagrams is shown in Fig.2. Such diagrams can be considered as the first terms
 in renormalization
series for PP line and can be reformulated with the help of "external electron line" (Fig.2b):
\begin{equation}
{\cal G}_{el}^{0\alpha\beta} (\omega ) =  \sum\limits_{j}{t_{ij} \,G_{jj}^{0\alpha\beta} (\omega )\,} t_{ji}
\end{equation}

\begin{figure}[h!]
\centerline{\includegraphics[width=6cm]{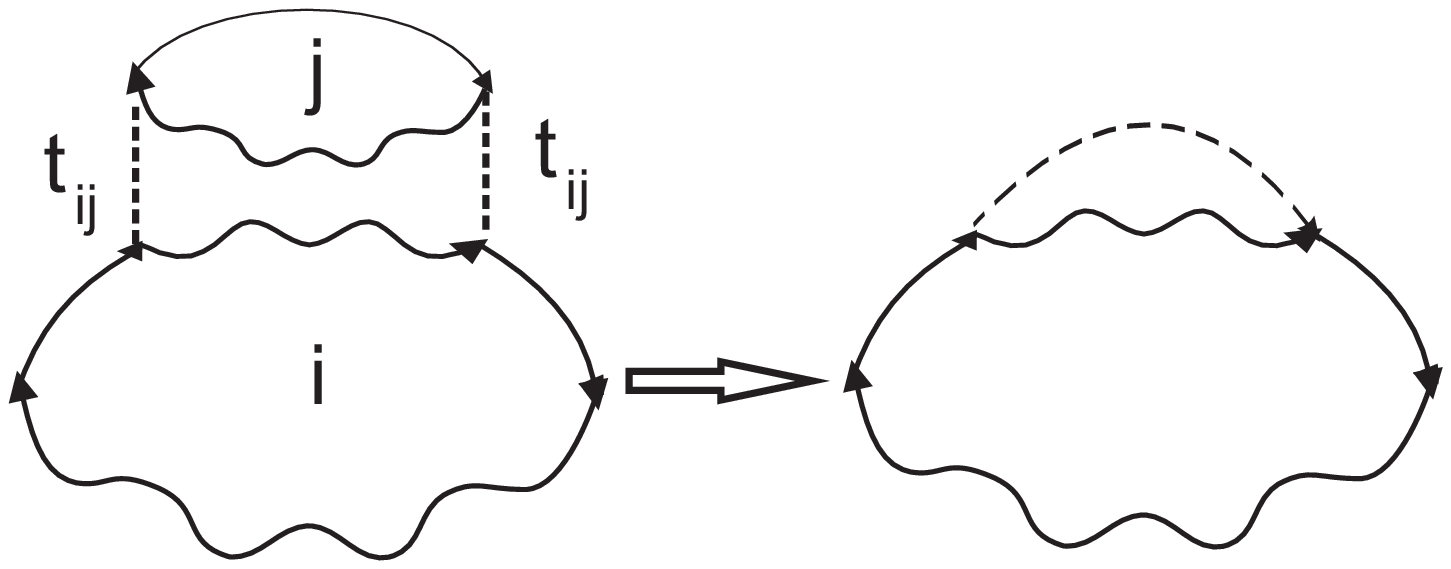}}
\caption{
a) Example of the first order diagram for electron on-site Green function. b) The same diagram in terms
of "external electron line"
${\cal G}_{el}^{0\alpha\beta} (\omega ) =  \sum\limits_{j}{t_{ij} \,G_{jj}^{0\alpha\beta} (\omega )\,} t_{ji}$
}
\end{figure}

If electron only once leaves a given site and returns back but we sum up all perturbation series for the
other sites, then this "external electron line" can be written in the same way
\begin{equation}
{\cal G}_{el}^{\alpha\beta} (\omega ) =  \sum\limits_{il}{t_{il} \,G_{lm}^{\alpha\beta} (\omega )\,} t_{mi}
\end{equation}
where $G_{lm}^{\alpha\beta}$ is the exact electron Green function for the problem with excluded given site $i$
(Fig 3).
\begin{figure}[h!]
\centerline{\includegraphics[width=6cm]{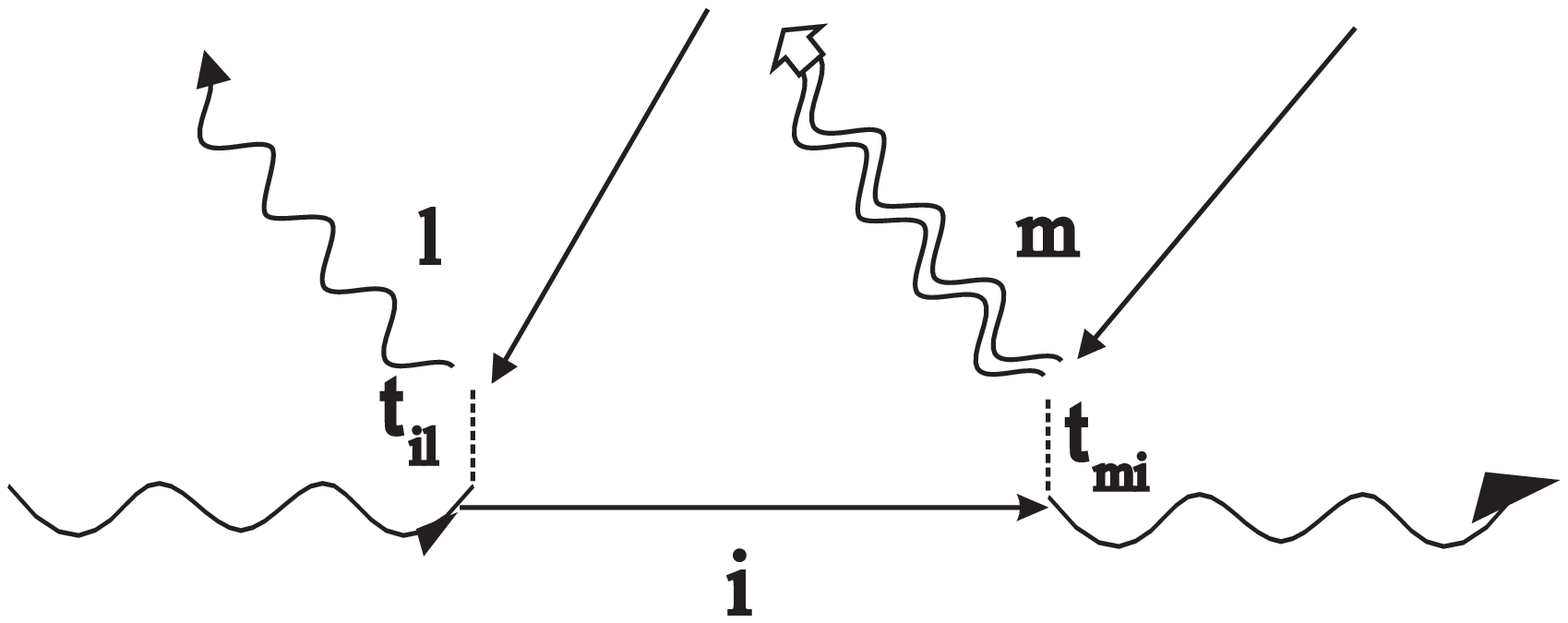}}
\caption{
Origin of self-consistent "external electron line"
${\cal G}_{el}^{\alpha\beta} (\omega ) =  \sum\limits_{il}{t_{il} \,G_{lm}^{\alpha\beta} (\omega )\,} t_{mi}$.
Hopping from site $i$, propagation in the surroundings and hopping back.
}
\end{figure}

Besides the diagrams which look like renormalization of PP lines there are also vertex corrections diagrams
(Fig 4). Calculations in the lowest orders show that vertex corrections are less important then diagrams
with renormalized PP lines, because they contain no secular divergences.

\begin{figure}[h!]
\centerline{\includegraphics[width=4.6cm]{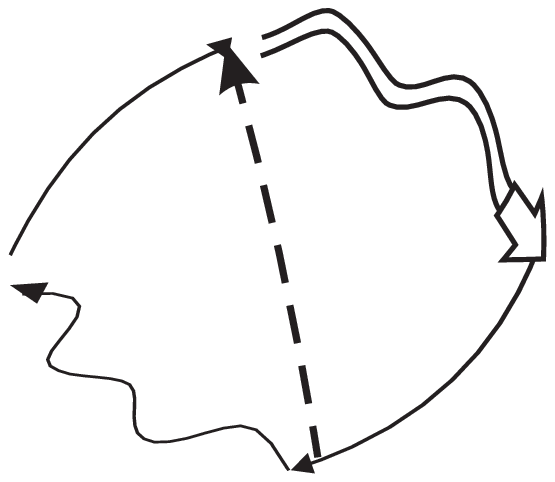}}
\caption{
Example of vertex corrections which are small compared to diagrams with renormalized PP lines
}
\end{figure}

So we propose self-consistent scheme for calculating electron Green functions based on
renormalization of PP lines only. Since in any PP loop only one PP occupation number is present the structure of
any diagram for on-site retarded electron Green function is strictly determined. Except the only one PP
occupation number all other PP lines are retarded or advanced functions ordered as it is
shown in Fig.5.

\begin{figure}[h!]
\centerline{\includegraphics[width=6cm]{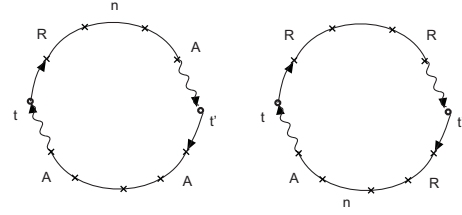}}
\caption{
The structure of
 diagrams for on-site retarded electron Green function. "$n$" denotes the only PP occupation number
 determined by Eqs.(\ref{n_0})
}
\end{figure}

The PP line, which contains PP ocupation number, will be called PP "lesser" Green function. And the other one, which
contains only retarded or advanced PP lines, will be called PP retarded or advanced Green functions.
Let us notice that from now on we use the term PP Green function for an object which is some diagram series
(upper or lower line in Fig.5) for which we can construct Dyson equation, but
strictly speaking sum of all diagrams for these lines are not usual particle Green functions.

Summing up all diagrams for PP retarded (advanced) Green function arising from diagrams like the first order
correction (Fig.2 ) we obtain Dyson equations in frequency representation:
\begin{eqnarray}
  \label{NR}
N^R_\sigma (\omega ) &=& N^{0R}_\sigma (\omega )+
N^{0R}_\sigma (\omega )\Sigma _{N\sigma}^R (\omega )N^R_\sigma(\omega ) \nonumber \\
B^R_\sigma (\omega ) &=& B^{0R} (\omega )+ B^{0R}(\omega )\Sigma _B^R (\omega )B^R(\omega )\\
D^R_\sigma (\omega ) &=& D^{0R}  (\omega )+ D^{0R} (\omega )\Sigma _D^R (\omega )D^R(\omega )\nonumber
\end{eqnarray}
where $N,B,D$ stand for the single-occupied site fermion, empty-site boson and double-occupied site
boson respectively. Zero order functions $N^0,B^0,D^0$ are given by Eq.(\ref{N_0} ).

We use an approximation in which self energy parts $\Sigma^R$ are determined by:
\begin{eqnarray}
  \label{SigmaR}
\Sigma _{N\sigma}^R (\omega ) &=& i\sum\limits_\sigma  \int {\frac{{d\omega _1 }}{{2\pi }}}
\left[ {\cal G}_{\sigma}^{el >}  (\omega _1 )B^R (\omega  - \omega _1 )\right. +  \nonumber \\
&+& \left.{\cal G}_{-\sigma}^{el <}  (\omega _1 )D^R (\omega  + \omega _1 ) \right] \nonumber \\
\Sigma _B^R (\omega ) &=& i\sum\limits_\sigma   \int {\frac{{d\omega _1 }}{{2\pi }}}
{\cal G}_{\sigma}^{el <}  (\omega _1 )N^R_\sigma (\omega  + \omega _1 )   \\
\Sigma _D^R (\omega ) &=& i\sum\limits_\sigma  \int {\frac{{d\omega _1 }}{{2\pi }}}
 {\cal G}_{-\sigma}^{el >}  (\omega _1 )N^R_\sigma (\omega  - \omega _1 ) \nonumber
\end{eqnarray}
where function ${\cal G}_{\sigma}^{<(>)}$ (eq.  ) in $k,\omega$ representation is:
\begin{equation}
     \label{Gcal}
{\cal G}_{\sigma}^{<(>)}(\omega ) = \sum\limits_k {\varepsilon _k^2 \,} G_{el\sigma}^{<(>)}(\omega ,k)
\end{equation}
Electron Green functions in equilibrium satisfy the following relations:
\begin{equation}
G_{el\sigma}^<  (\omega ,k) =  - 2i\,f(\omega ){\rm Im} G_{el\sigma}^R (\omega ,k)
\end{equation}
\begin{equation}
G_{el\sigma}^ >  (\omega ,k) =  - 2i\,(f(\omega )-1){\mathop{\rm Im}\nolimits} G_{el\sigma}^R (\omega ,k)
\end{equation}
And retarded electron Green function $G_{el\sigma}^R$ should be determined later self-consistently.

\begin{figure}[h!]
\centerline{\includegraphics[width=8cm]{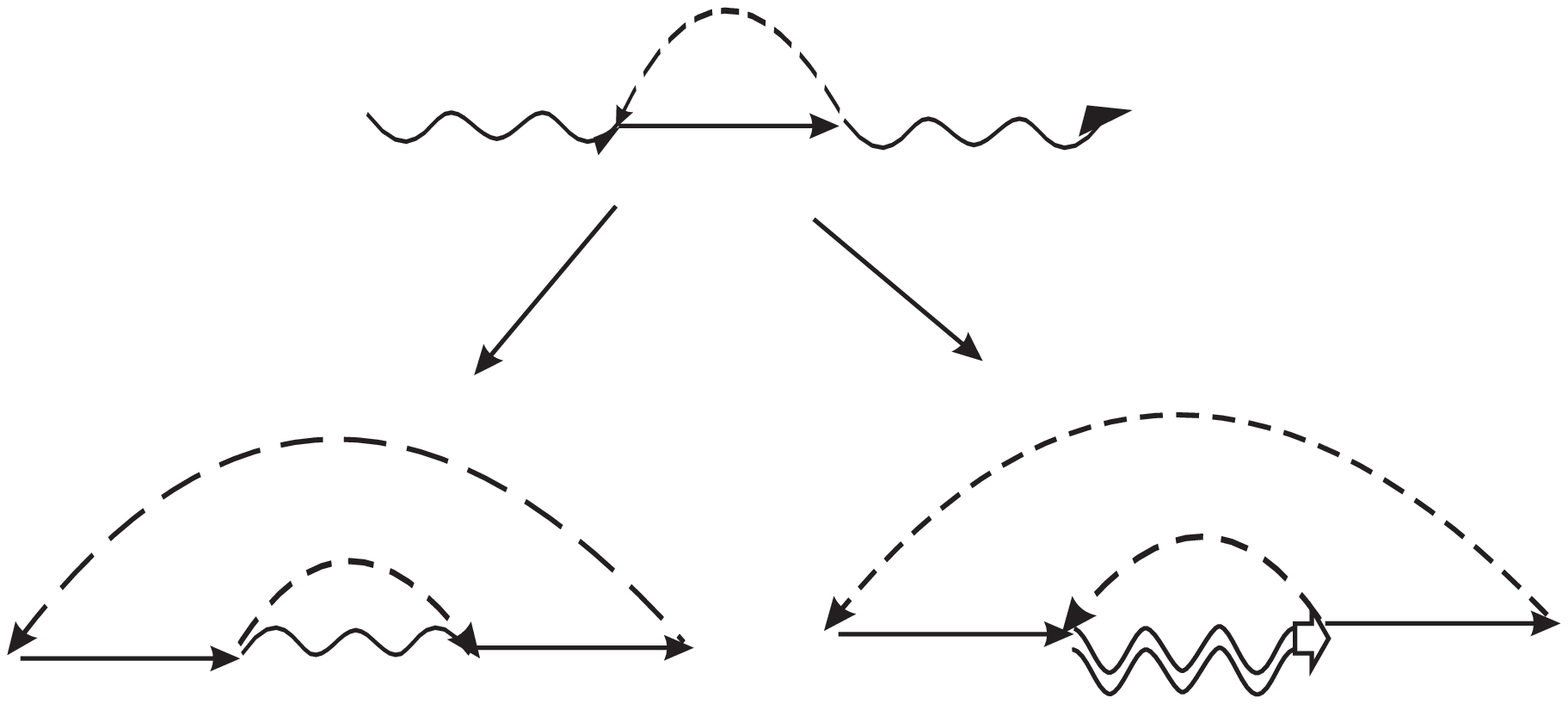}}
\caption{
Example of self-energy parts used in the suggested approximation.
Dashed line - "external electron line" Eq.(\ref{Gcal})
 }
\end{figure}

Dyson equations with self energy parts (\ref{SigmaR}) mean that we sum up diagram series of the type shown
in Fig.6. Of course this is an approximation because we take into consideration interaction on
a given site but replace complicated correlated electron transport through
all other sites by a sum of uncorrelated processes (hopping from the site, propagation in the surroundings
and hopping back). This propagation is described by some averaged single electron function which
in self-consistent procedure should be determined by means of the same electron function calculated
for our given site.
The idea in some sense resembles the dynamical mean field theory (DMFT) though the approach itself and
all basic equations are quite different.

Let us point out that retarded (advanced) PP self energies contains no PP occupation numbers, thus
these self energies are incomplete compared to the case of real particles: the part with "lesser" PP functions
is omitted. So, in spite of the Dyson equation for this diagram series has the usual form, it can not
be regarded as an equation for some real particle Green function.
If we know $G_{el\sigma}^R$ the system of equations (\ref{NR},\ref{SigmaR}) is complete and allows to calculate
self-consistently all PP retarded or advanced functions.

Dyson equation for PP "lesser" Green function (lines with one occupation PP number) can be written in a
similar way:
\small
\begin{eqnarray}
   \label{N<}
N^<_\sigma(\omega )& =& N^R_\sigma (\omega )\left[\Sigma _{N\sigma}^<(\omega )
+(N_0^R)^{-1}N^{0<}_\sigma (\omega )(N_0^A)^{-1}\right] N^A_\sigma (\omega ) \nonumber\\
B^ <  (\omega )& =& B^R (\omega )\left[ \Sigma _B^ <  (\omega )
+(B_0^R)^{-1}B^{0<}(\omega )(B_0^A)^{-1}\right] B^A (\omega ) \nonumber \\
D^ <  (\omega ) & =& D^R (\omega )\left[\Sigma _D^ <  (\omega )
+(D_0^R)^{-1}D^{0<}(\omega )(D_0^A)^{-1}\right] D^A (\omega )   \nonumber \\
\end{eqnarray}
\normalsize
And "lesser" self energy parts look like:
\begin{equation}
  \label{Sigma<}
\begin{array}{l}
\Sigma _{N\sigma}^ <  (\omega ) = \nonumber \\
 i \int {\frac{{d\omega _1 }}{{2\pi }}}
\left[ {{\cal G}_{\sigma}^{el <}  (\omega _1 )B^ <  (\omega  - \omega _1 ) +
{\cal G}_{-\sigma}^{el >}  (\omega _1 )D^ <  (\omega  + \omega _1 )} \right] \nonumber \\
\Sigma _B^ <  (\omega ) = i\sum\limits_\sigma \int {\frac{{d\omega _1 }}{{2\pi }}}
{\cal G}_{\sigma}^{el >} (\omega _1 )N^ <  (\omega  + \omega _1 )                      \\
\Sigma _D^ <  (\omega ) = i\sum\limits_\sigma\int {\frac{{d\omega _1 }}{{2\pi }}}
{\cal G}_{-\sigma}^{el <}  (\omega _1 )N^<_\sigma  (\omega  - \omega _1 ) \nonumber
\end{array}
\end{equation}
The system of equations (\ref{N<},\ref{Sigma<}) is also complete since PP retarded and advanced functions
have been calculated already. So all PP lesser functions can be determined
self-consistently from these equations.

Now we can calculate the on-site electron Green function from the same diagrams as in Fig.1 but with
renormalized ("dressed") PP Green functions:
\begin{equation}
   \label{GRel}
\begin{array}{l}
 G_{ii\sigma}^{R\,el} (\omega ) = \\
 i\int {\frac{{d\omega _1 }}{{2\pi }}}
 \left[ {B_{}^<(\omega _1 )N^R_\sigma(\omega  + \omega _1 )+ B_{}^A (\omega _1 )N^<_\sigma(\omega  + \omega _1 )} \right. -  \\
 \left.  - N_{ - \sigma }^ <  (\omega _1 )D^R (\omega  + \omega _1 )
 - N_{ - \sigma }^A (\omega _1 )D^ <  (\omega  + \omega _1 ) \right]
 \end{array}
\end{equation}
In this paper within the simplest approximation we shall consider the usual relation
between on-site and band electron Green functions:
\begin{equation}
   \label{Gk}
G_{el\sigma}^R (\omega ,k) = \frac{1}{{(G_{ii\sigma}^{R\,el})^{-1}(\omega )  - \varepsilon _k }}
\end{equation}
Since all PP functions in Eq.(\ref{GRel}) can be calculated if we know electron Green function
$G_{el\sigma}^R (\omega ,k)$ , we get to a self-consistent scheme of calculations. The steps are the following:

1) From zero-order on-site Green function Eq.(\ref{G0}) we calculate electron Green function Eq.(\ref{Gk})
 and "external line electron function" Eq.(\ref{Gcal}) for PP diagrams.

2)Perform self-consistent calculations first of R,A, and then of "lesser" PP functions.

3)Determine new on-site electron Green function from Eq.(\ref{GRel}) and proceed with new band electron
(\ref{Gk}) and
"external line electron" (\ref{Gcal}) functions.

This procedure should be repeated until the stable solution is reached.

In these calculations we encounter with some differences from usual calculations with real particle Green
functions. The functions which are called PP Green functions are just some diagram subseries so their
properties are not obligatory the same as for the real particle Green functions.
Their spectral weight is not automatically normalized for example. Our scheme of calculations gives us the
shape of electron density of state but not its absolute value. So we require that retarded
electron Green function should be normalized as usual
$$
-\frac{1}{\pi}\int d\omega \sum\limits_k {\rm Im} G_{el\sigma}^R (\omega ,k)=1
$$

\subsection*{Results}
The shape of the electron density of states  depends on the value of Coulomb interaction and
electron concentration. We present here results of calculations for 2D square lattice.
For the half filling case ($\mu=\varepsilon+U/2$) we see that two Hubbard subbands with dielectric
gap between them begin to form for Coulomb repulsion comparable with the bandwidth. With further increasing
of Coulomb repulsion the two-subband structure with well defined gap is more and more pronounced (Fig.7).
Let us stress that there is no artifact central peak which usually appears in DMFT calculations.
Van-Hove singularity of 2D noninteracting electron band is completely smoothed for large enough $U$ due to
interaction.

\begin{figure}[h!]
\includegraphics[width=8.5cm]{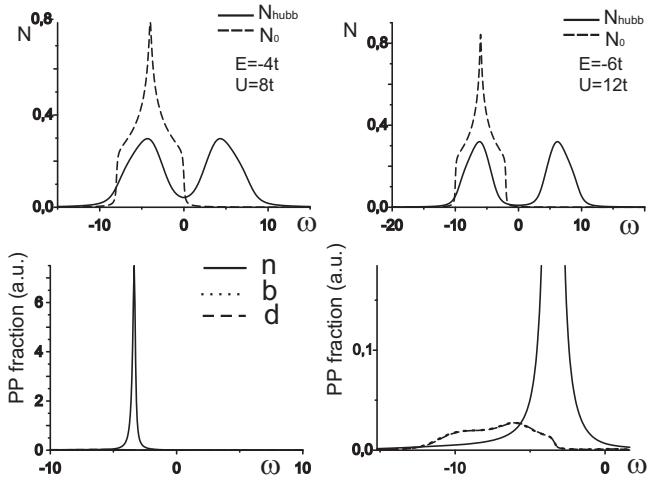}
\caption{
Electron density of states (solid line) for the half filling case and  different $U$ (upper panels):
density of states for the band with no interaction is shown by the dashed line, $U$ is measured in units of
intersite hopping $t_{ij}$. In the lower raw PP "lesser" functions are depicted for $U=12t$. They show the
relative fraction of empty (dotted line), single-occupied (solid line) and double occupied (dashed line) sites.
The right figure - enlarged part of the left one to demonstrate the small admixture of empty and
double-occupied sites.
}
\end{figure}

If the band is almost empty the two-subband structure is nearly destroyed and finally the density of states
for noninteracting electron band is restored (Fig.8a). Similar picture is observed for
almost filled band (Fig.8b).
The the density of states for noninteracting electron band is again restored
but it is shifted up in energy by the value of the Coulomb interaction. In  both these cases Van-Hove
singularity of 2D noninteracting electron band appears in the density of states.
\begin{figure}[h!]
\includegraphics[width=8.5cm]{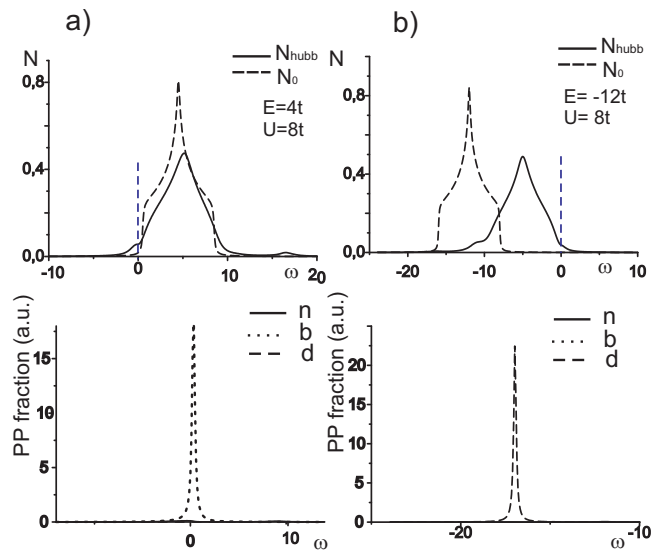}
\caption{
Electron density of states (solid line) for almost empty (a) and almost filled (b) band (upper panels).
Vertical line is the position of the chemical potential.
Lower raw - corresponding relative fractions of empty (dotted line), single-occupied (solid line)
and double occupied (dashed line) sites.
}
\end{figure}
Presented scheme allows to find density of states for any intermediate electron concentration. The modification
of two-band Hubbard structure with concentration changes is shown in Fig.9 .
\begin{figure}[h!]
\includegraphics[width=8.5cm]{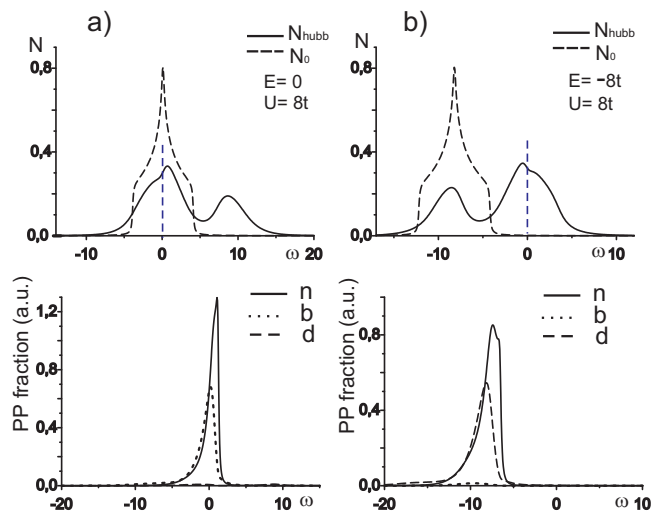}
\caption{
Electron density of states (solid line) for intermediate concentrations (upper panels): (a) - concentration
is less and (b) - greater than half filling. Lower raw -  corresponding relative fractions of differently
occupied sites. The notations are the same as in Figs.7,8.
}
\end{figure}

Though "lesser" PP functions are not Green functions of real particles, nevertheless their relative values
reflect ratio between the numbers of  empty, single-occupied and double -occupied sites. We see that these
functions quite reasonably describe the physical situation for different electron
concentrations and Coulomb repulsion values. It is very important that self-consistent solution for "lesser" PP
functions are independent on initial PP occupation numbers $n_0,b_0,d_0$ (as it should be in non-perturbative
calculations in Keldysh technique).
At half-filling most sites are single-occupied as it is evident from Fig.7 : "lesser" function of
pseudo-fermion dominates and admixture of two pseudo-bosons for empty and double-occupied sites is negligible.
For almost empty (or almost filled) band vice versa the weight of "empty"-pseudo-boson  (or
double-occupied-boson) is the largest one and fraction of single-occupied sites is small.
When the concentration of electrons increases from empty to completely filled band
we can analyze how the fraction of differently occupied sites changes (Fig.9).

The external electron line ${\cal G}_{\sigma}(\omega ) $ is similar to electron self energy
for on-site Green function in Hubbard-III
approximation \cite{Hubb}. But in Hubbard-III approximation the self energy  is multiplied by initial (fixed)
occupation electron numbers. The present approach takes into account self consistent changes of
electron on-site occupation due to Coulomb interaction via PP "lesser" functions calculations.

We should mention that this approach works well if parameter $t^2/(\varepsilon +U)^2 $ or
$t^2/\varepsilon^2 $ is less than unity. So this simple approximation can not give correct
result for the half filling situation and small $U$ when both parameters become greater than unity.

\subsection*{Conclusions}
We suggested a new approach to describe properties of correlated electron systems based on pseudo-particle
Keldysh diagram technique.  For the first time consistent non-perturbative calculations
in pseudo-particle technique was performed. For the Hubbard model pseudo-particle technique can
give reasonable results for electron density of states for different electron concentration and
Coulomb repulsion values. Note that DMFT can be hardly applied to arbitrary concentrations different from
the half filling case. Even at half-filling DMFT always gives artificial central peak in the density of states
which does not appear in our method.

It is possible to calculate self consistently "lesser" functions for pseudo particles which are
independent on their  initial occupation numbers. These "lesser" functions quite reasonably reproduce
the relative fractions of empty, single- and double- occupied sites.

The main advantages of this approach are that it allows to work in real time representation and
does not need analytical  continuation as in temperature diagram technique.
Keldysh technique can be applied for any temperatures  as well as for nonequilibrium and even
nonstationary situation. At last compared to DMFT our  calculations are very fast.

This research was supported by RFBR grants 03-02-16807 and RAS Program "Strongly correlated electrons
in metals, semiconductors and superconductors".


\end{document}